\documentclass[longbibliography,aip,apl,reprint,groupedaddress,superscriptaddress]{revtex4-1} 

\setcitestyle{super,open={},close={}}
\usepackage{graphicx}
\graphicspath{{./figure_apl/}}
\usepackage{bm}
\usepackage{amsmath,amssymb}
\usepackage{hyperref}
\usepackage{lipsum}
\usepackage{color}
\definecolor{colorref}{rgb}{0.0, 0.408, 0.647}
\definecolor{grey}{rgb}{0.9, 0.9, 0.9}
\hypersetup{
	colorlinks = true,
	allcolors = colorref
}
\newcommand{\figtitle}[1]{{#1}}
\newcommand{\subfiglabel}[1]{(#1)}
\newcommand{\rsubfiglabel}[1]{({#1})}

\newcommand{\rfig}[1]{\textcolor{colorref}{Fig.~\ref{#1}}}

\newcommand{\rsubfig}[2]{\textcolor{colorref}{Fig.~\ref{#1}(#2)}}

\newcommand{\rsubfigsA}[3]{\textcolor{colorref}{Figs.~\ref{#1}(#2)} and \textcolor{colorref}{\ref{#1}(#3)}}

\definecolor{reviseColor}{rgb}{0, 0.0, 0.0}
\newcommand{\revise}[1]{\textcolor{reviseColor}{#1}}

\newcommand{\subpart}[1]{\noindent\textbf{#1}}
\newcommand{\SeeSupply}[1]{Supplemental Material}

\newcommand*{\citen}[1]{%
  \begingroup
    \romannumeral-`\x 
    \cite{#1}%
  \endgroup
}

\usepackage[normalem]{ulem}  

\newcommand{\SIQSE}{\affiliation{1}{Shenzhen Institute for Quantum Science and Engineering, Southern University of Science and Technology, Shenzhen, Guangdong, China}}

\newcommand{\IQA}{\affiliation{3}{International Quantum Academy, Shenzhen, Guangdong, China}}
\newcommand{\GDKL}{\affiliation{4}{Guangdong Provincial Key Laboratory of Quantum Science and Engineering, Southern University of Science and Technology, Shenzhen, Guangdong, China}}
\newcommand{\HFNL}{\affiliation{5}{
Shenzhen Branch, Hefei National Laboratory, Shenzhen 518048, China}}
\newcommand{\NXU}{\affiliation{6}{
School of Physics and Electronic-Electrical Engineering, Ningxia University, Yinchuan, 750021, China}}

\begin{document}

\title{Dephasing-assisted diffusive dynamics in superconducting quantum circuits}

\date{\today}

\author{Yongqi Liang}
\thanks{These authors contributed equally to this work.}
\affiliation{\SIQSE}\affiliation{\IQA}\affiliation{\GDKL}
\author{Changrong Xie}
\thanks{These authors contributed equally to this work.}
\affiliation{\SIQSE}\affiliation{\IQA}\affiliation{\GDKL}

\author{Zechen Guo}
\affiliation{\SIQSE}\affiliation{\IQA}\affiliation{\GDKL}

\author{Peisheng Huang}
\affiliation{\NXU}\affiliation{\IQA}

\author{Wenhui Huang}
\affiliation{\SIQSE}\affiliation{\IQA}\affiliation{\GDKL}

\author{Yiting Liu}
\affiliation{\SIQSE}\affiliation{\IQA}\affiliation{\GDKL}

\author{Jiawei Qiu}
\affiliation{\SIQSE}\affiliation{\IQA}\affiliation{\GDKL}

\author{Xuandong Sun}
\affiliation{\SIQSE}\affiliation{\IQA}\affiliation{\GDKL}

\author{Zilin Wang}
\affiliation{\NXU}\affiliation{\IQA}

\author{Xiaohan Yang}
\affiliation{\SIQSE}\affiliation{\IQA}\affiliation{\GDKL}

\author{Jiawei Zhang}
\affiliation{\SIQSE}\affiliation{\IQA}\affiliation{\GDKL}

\author{Jiajian Zhang}
\affiliation{\SIQSE}\affiliation{\IQA}\affiliation{\GDKL}

\author{Libo Zhang}
\affiliation{\SIQSE}\affiliation{\IQA}\affiliation{\GDKL}

\author{Ji Chu}
\affiliation{\IQA}

\author{Weijie Guo}
\affiliation{\IQA}

\author{Ji Jiang}
\affiliation{\SIQSE}\affiliation{\IQA}\affiliation{\GDKL}

\author{Xiayu Linpeng}
\affiliation{\IQA}

\author{Song Liu}
\affiliation{\SIQSE}\affiliation{\IQA}\affiliation{\GDKL}\affiliation{\HFNL}

\author{Jingjing Niu}
\affiliation{\IQA}\affiliation{\HFNL}

\author{Yuxuan Zhou}
\affiliation{\IQA}

\author{Youpeng Zhong}
\email{zhongyp@sustech.edu.cn}
\affiliation{\SIQSE}\affiliation{\IQA}\affiliation{\GDKL}\affiliation{\HFNL}

\author{Wenhui Ren}
\email{wenhuiren@zju.edu.cn}
\affiliation{\IQA}

\author{Ziyu Tao}
\email{taoziyu@iqasz.cn}
\affiliation{\IQA}

\author{Dapeng Yu}
\affiliation{\SIQSE}\affiliation{\IQA}\affiliation{\GDKL}\affiliation{\HFNL}

\date{\today}

\begin{abstract}
Random fluctuations caused by environmental noise can lead to decoherence in quantum systems.
Exploring and controlling such dissipative processes is both fundamentally intriguing and essential for harnessing quantum systems to gain practical advantages and deeper insights.
In this work, we first demonstrate the diffusive dynamics assisted by controlled dephasing noise in superconducting quantum circuits, contrasting with coherent evolution.
We show that dephasing can \revise{give distinct dynamical behavior in a superconducting qubit array with quasiperiodic order.}
Furthermore, by preparing different excitation distributions in the qubit array, we observe that a more localized initial state relaxes to a uniformly distributed mixed state faster with dephasing noise,
illustrating another counterintuitive phenomenon called \revise{Mpemba-effect-like quantum dynamics}, i.e., a far-from-equilibrium state can relax toward the equilibrium faster.
These results deepen our understanding of diffusive dynamics at the microscopic level, and demonstrate controlled dissipative processes as a valuable tool for investigating Markovian open quantum systems.
\end{abstract}
\maketitle

Decoherence is an inherent feature of quantum systems, arising from the inevitable interactions with the numerous degrees of freedom in the surrounding environment.
Such random interactions generally lead to two types of dissipative processes in quantum systems, i.e., energy relaxation caused by spontaneous emission, and dephasing caused by random fluctuations in the system parameters.
Exploring and controlling these dissipative processes is not only interesting in itself from fundamental aspects, but also crucial for harnessing quantum systems to gain any technical advantage or insight~\cite{Harrington2022}.
Uncontrolled dissipative processes pose a major challenge in quantum information processing, as they erode essential quantum resources like coherence and entanglement, degrade the fidelity of quantum gates, and add noise to measurement signals.
However, engineered dissipative processes can play a constructive role, for example, in
quantum measurement~\cite{RevModPhys.82.1155,ChenHongzhen2024a},
quantum state stabilization~\cite{Lin2013,Leghtas2015,Lu2017,Grimm2020,Lingenfelter2024},
autonomous quantum error correction~\cite{Gertler2021,Xu2023,Li2024},
dissipative phase transition~\cite{Prosen2008,Kessler2012,Roberts2023,Roberts2023a},
long-range coherence~\cite{Dutta2020,Li2025},
and non-Hermitian systems~\cite{Bergholtz2021}.

\begin{figure}[!t]
	\centering
    \includegraphics[width=0.46\textwidth]{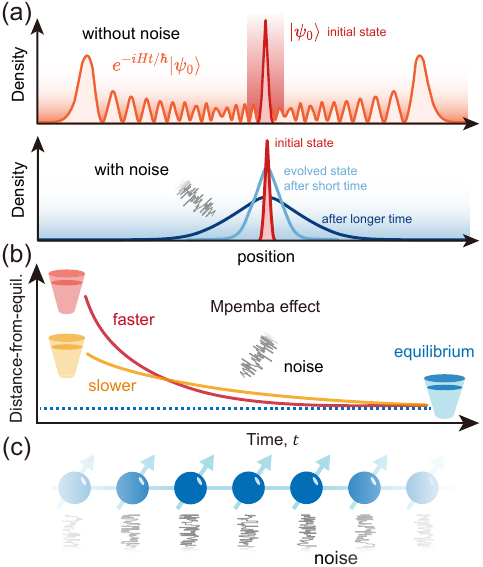}
	\caption{\label{fig1}
    Dephasing-assisted transition from ballistic to diffusive transport and accelerated equilibration in a 1D qubit array.
    \subfiglabel{a}
    Illustration of the density profile for an initial state $\vert \psi_0\rangle$ evolved under Hamiltonian $H$ without (upper panel) and with dephasing noise (lower panel).
    \subfiglabel{b}
    \revise{Schematic for Mpemba-effect-like dynamics}, where a far-from-equilibrium state (red line) reaches equilibrium (denoted as blue dashed line) faster than a state closer to equilibrium (yellow line) under certain conditions (here with dephasing noise).
    \subfiglabel{c}
    The 1D array model used in this experiment to demonstrate the above phenomena, where controlled dephasing is incorporated by randomly modulating the qubit frequencies on each site.
}
\end{figure}

While dephasing noise generally spoils quantum correlations, it can lead to collective phenomena, including
noise-induced transport~\cite{Lindner2002,Gopalakrishnan2017,Ren2020},
quantum synchronization~\cite{Schmolke2022,Tao2025}, and quantum phase transition~\cite{LiuGGC2018}.
In quantum many-body systems, dephasing noise acts as a dynamically fluctuating disorder, which is generally acknowledged to enhance transport and spoil \revise{localized dynamical behavior}~\cite{Gurvitz2000,Plenio2008,Levi2012,Nakajima2018}.
Recently, this common wisdom has been theoretically revisited, finding that \revise{the dynamical behavior can be changed} by dephasing noise in one-dimensional (1D) quasiperiodic systems, in the regime where all the eigenstates are spatially extended in the coherent counterpart~\citen{Longhi2024}, as illustrated in \rsubfig{fig1}{a}.
The excitation in the lattice spreads via a diffusive process assisted by the dephasing effect, showing distinct \revise{dynamical behavior} in contrast with the ballistic dynamics under coherent evolution.
Dephasing noise can also assist the illustration of another counterintuitive phenomenon called \revise{Mpemba-effect-like quantum dynamics}~\cite{Longhi2024a}, which means a far-from-equilibrium state can relax toward equilibrium faster~\cite{Bechhoefer2021}, as illustrated in \rsubfig{fig1}{b}.
These theoretical findings provided rigorous insights into the quantum dynamics besides coherent evolution~\cite{Carollo2021}, appealing for further experimental explorations.

Featuring scalability and controllability, superconducting quantum circuits are not only one of the leading candidates for quantum computation~\cite{Google2023}, but also an ideal playground for exploring exotic phenomena with controlled dissipation.
For example, superconducting qubits have been used to explore dissipative many-body dynamics~\cite{Ma2019,Mi2024,Du2024},
non-Hermitian physics~\cite{Naghiloo2019,Chen2021,Chen2022},
and quantum state stabilization~\cite{Touzard2018,Li2024b,Brown2022,Chen2025,LiSai2024a}.
In this work, we demonstrate the dephasing-assisted diffusive dynamics and associated phenomena using a 1D array of seven superconducting qubits with tunable nearest-neighbor couplings, as illustrated in \rsubfig{fig1}{c}.
The dynamics of this system are captured by the effective Hamiltonian of a 1D tight-binding model with $L=2l+1$ sites:
\begin{equation}
	H/\hbar = \sum_{j=-l}^{l} \Delta \omega_j \sigma_j^{+} \sigma_j^{-}  + 
    \sum_{j=-l+1}^{l}  g_j (\sigma_j^{+} \sigma_{j-1}^{-} + \sigma_j^{-} \sigma_{j-1}^{+})
	,\label{eq:Hami1d}
\end{equation}
where $\sigma^{+}_{j}$ ($\sigma_{j}^{-}$) represents the raising (lowering) operator for the qubit at site $j$,
$\sigma_j^{+} \vert 0\rangle^{\otimes L} = \vert 1_j\rangle $,
$\Delta \omega_j/2\pi$ is the frequency detuning of the qubit relative to the average system frequency,
$g_j/2\pi$ is the tunable coupling strength between site $j-1$ and $j$, and $l=3$ here, see \SeeSupply{} for more details of the device.
By adjusting the amplitude of microwave pulses applied to the qubits, we can tune the qubits on resonance with $\Delta \omega_j = 0$.
To introduce controlled dephasing noise into the system, we apply a series of white noises $\xi_j(t)$ to the modulation of the qubit frequency $\Delta\omega_j$.
The evolution of density matrix $\rho$ can be effectively described by the Markovian dynamics generated in the Lindblad master equation:
\begin{equation}
	\frac{d\rho}{dt} = -\frac{i}{\hbar} \left[H, \rho\right] + 
	\sum_{j} \left(\mathcal{L}_j \rho \mathcal{L}_j^\dagger - \frac{1}{2} \left\{\mathcal{L}_j^\dagger \mathcal{L}_j ,\rho \right\} \right),
\end{equation}
where $\mathcal{L}_j=\sqrt{\Gamma_j}\sigma_{j}^{+}\sigma_{j}^{-}$ is the collapse operator for dephasing,
$\Gamma_j$ is the effective dephasing rate introduced by the noise $\xi_j(t)$ with $\Gamma_j=\Gamma$, see \SeeSupply{} for details.

\begin{figure}[!t]
    \centering
    \includegraphics[width=0.46\textwidth]{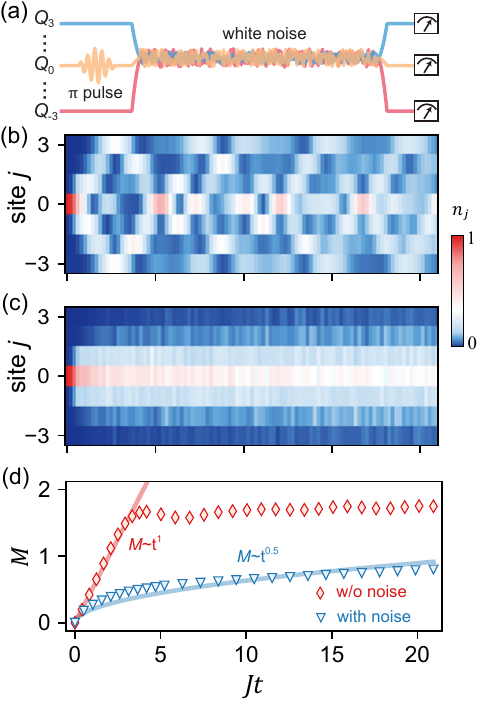}
    \caption{\label{fig2}
    \figtitle{Diffusive spreading dynamics with dephasing noise.}
    \subfiglabel{a}
    Schematic of the experimental pulse sequence.
    \subfiglabel{b}-\subfiglabel{c}
    Measured population dynamics $n_j$ without \rsubfiglabel{b} and with \rsubfiglabel{c} dephasing noise,
    illustrating ballistic- and diffusive-like spreading, respectively.
    \subfiglabel{d}
    The integrated moment $M(t)$ as a function of normalized time $Jt$, 
    which captures the ballistic \rsubfiglabel{b} and diffusive manner of spreading \rsubfiglabel{c} during early dynamics.
    Dots indicate measured results, while the red (blue) solid line denotes the scaling function of $M\sim t^1$ ($t^{0.5}$).
    }
\end{figure}

\begin{figure*}[!t]
    \centering
    \includegraphics[width=0.95\linewidth]{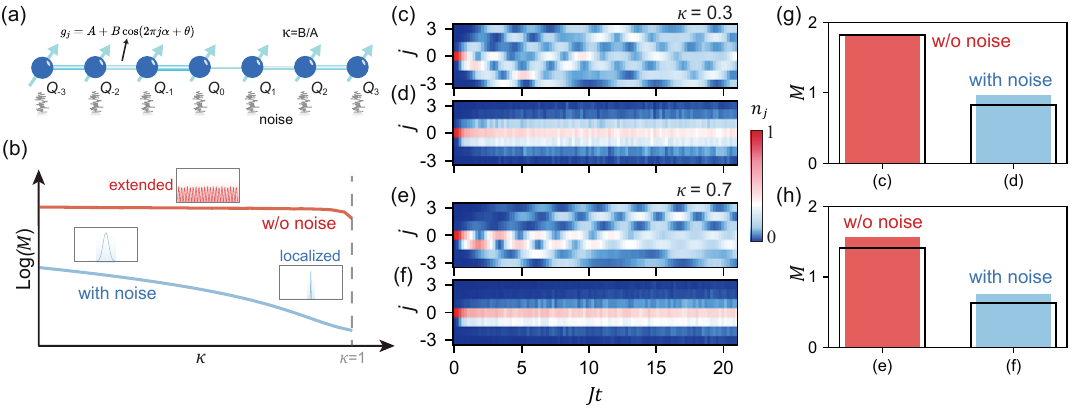}
    \caption{\label{fig3}
    \figtitle{\revise{Dephasing-assisted dynamics} in a quasiperiodic lattice.}
    \subfiglabel{a}
    Schematic for the off-diagonal Aubry-Andr\'e model with aperiodic coupling strengths, where the controlled dephasing noises are injected into the qubits.
    \subfiglabel{b}
    Numerical calculation of $M$ versus $\kappa$ with and without dephasing noise for $L=89$ sites.
    \subfiglabel{c}-\subfiglabel{f}
    Measured population dynamics $n_j(t)$ without \rsubfiglabel{c,e} and with \rsubfiglabel{d,f} noise for $L=7$ qubits,
    showing oscillatory dynamics and diffusive spreading behavior, respectively.
    \subfiglabel{g}-\subfiglabel{h}
    The integrated moment $M$ evaluated from the measured dynamics shown in \rsubfiglabel{c-f},
    where the solid bars and black frames denote the measured and numerical results.
    The ratio $\kappa = B/A$ is $0.3$ for \rsubfiglabel{c,d,g}, and $0.7$ for \rsubfiglabel{e,f,h}.
    }
\end{figure*}

We first demonstrate the transition from ballistic- to diffusive-like dynamics by investigating the qubit evolution.
In \rfig{fig2}, we present the temporal evolution of the measured on-site population 
$n_j(t) =\langle 1_j \vert \rho(t)\vert 1_j \rangle$ of individual qubits versus normalized evolution time $Jt$.
The coupling strengths $g_j = J \approx 2\pi\times 8.3~\mathrm{MHz}$ are homogeneous,
and the effective reduced dephasing rate $\Gamma/J \approx 30 \gg1$.
We examine the population dynamics of the qubit array by initially preparing the central qubit in its excited state $\vert 1_{j=0}\rangle$ using a $\pi$ pulse (see \rsubfig{fig2}{a}).
In the absence of noise, we should observe the excitation spreading in a ballistic manner throughout the lattice. This ballistic-like spreading is confirmed in \rsubfig{fig2}{b}, where the population extends rapidly across the qubits, with reflections at open boundaries due to the finite system size.
When noise is applied, the dephasing effect leads to a slowdown in spreading dynamics, transforming its behavior from ballistic to diffusive, as shown in \rsubfig{fig2}{c}.
To further analyze the spreading dynamics quantitatively, we use the integrated moment $M(t)=(1/t)\int_{0}^{t}[W(\tau)-W(0)]^{1/2}d\tau$ with the effective second moment $W(\tau)=\sum_{j}{|j-j_{0}|^2} n_{j}(\tau)$ and initial center $j_0=0$ as shown in \rsubfig{fig2}{d},
which captures the expansion of population dynamics from the initial center, giving
the characteristic feature of $M(t)\sim t^1$ for ballistic spreading and $M(t)\sim t^{0.5}$ for diffusive spreading with strong dephasing noise.
The saturated values of $M$ after $Jt\approx 4$ in the absence of noise are captured from the reflections of qubit population at open boundaries due to the finite system size shown in \rsubfig{fig2}{b}.

Recently, the off-diagonal Aubry-Andr\'e model~\cite{Aubry1980,Cestari2016,Xiao2021} has been revisited with dephasing noise~\cite{Longhi2024}, finding that the spreading dynamics of excitation can exhibit a different \revise{dynamical behavior} from the coherent evolution.
We consider this model shown in \rsubfig{fig3}{a}, where the incommensurate aperiodic coupling strength is introduced as
$g_j =A+B\cos(2\pi \alpha j+\theta)$, with $A,B >0$, irrational frequency $\alpha = \left(\sqrt{5}-1 \right)/2$, and phase offset $\theta=0$ throughout this work.
In the regime of $\kappa = B/A <1$ for the off-diagonal Aubry-Andr\'e model, all the eigenstates are spatially extended~\cite{Xiao2021}.
In \rsubfig{fig3}{b}, we numerically show the behavior of the integrated moment $M$ for different values of $\kappa$, both with and without dephasing noise, where $L=89$ is a Fibonacci number used to numerically evaluate the model with a large system size,
\revise{and the time scale from $Jt=0$ to $Jt=50$ is chosen sufficiently large for numerically evaluating the integrated moment $M$ with the system size $L=89$}.
It is evident that $M$ without dephasing noise is an order of magnitude larger than that with noise.
To experimentally investigate the \revise{quantum dynamics}, we focus on the early time dynamics of a small finite system tailored to the superconducting quantum circuits, in which \revise{the dynamical behavior} can still be distinguished from the experimental observables~\cite{Xiao2021,Bravyi2006}.

Here, we present the experimental demonstration for \revise{dephasing-assisted dynamics} described by the off-diagonal Aubry-Andr\'e model with dephasing noise.
By configuring the coupling strength as quasiperiodic order through the tunable couplers between qubits, we investigate such intriguing phenomena by the characteristic dynamics of qubit excitation.
We first explore the coherent evolution without artificial dephasing noise, where $A+B = J \approx 2\pi \times 8.3~\mathrm{MHz}$, and different $\kappa$ within the regime of extended phase.
We initialize a single-particle excitation in the center site, and observe the population of each site versus normalized time $Jt$ as shown in \rsubfigsA{fig3}{c}{e} for $\kappa=0.3$ and 0.7, respectively, where the dynamics show clear oscillations traversing all sites, consistent with the picture of extended phase.
When the applied noises introduce an intense dephasing noise with $\Gamma/J\approx 30$ on the lattice system, localized eigenstates with an extremely long lifetime emerge,
which strongly influences the dynamic spreading of excitation and slows down the spreading dynamics~\cite{Longhi2024}.
The measured dynamics with dephasing noise for $\kappa=0.3$ and 0.7 are shown in \rsubfigsA{fig3}{d}{f} respectively, showing a significantly slowing down diffusive spreading dynamics as compared to their counterpart without noise.
The slowing down of diffusive spreading is also captured by the integrated moment $M$ accumulated during the evolution \revise{from $t=0~\mathrm{ns}$ to $t=400~\mathrm{ns}$} in \rsubfigsA{fig3}{d}{f}, as shown in \rsubfigsA{fig3}{g}{h}.

\begin{figure}[!t]
    \centering
    \includegraphics[width=0.46\textwidth]{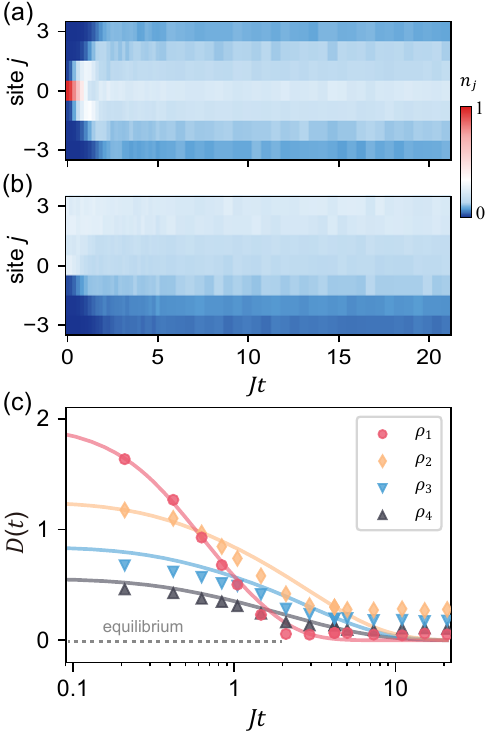}
    \caption{\label{fig4}
    \figtitle{\revise{Mpemba-effect-like dynamics with dephasing noise.}}
    \subfiglabel{a}-\subfiglabel{b}
    Measured population dynamics $n_j$ with a dephasing rate $\Gamma/J\approx 3$, where the system is initially prepared in states with density matrices 
    \subfiglabel{a} $\rho_1 = \vert 1_0 \rangle \langle 1_0 \vert$,
    and \subfiglabel{b} $\rho_4 = 1/4\sum_{j=0}^{3} \vert 1_j \rangle \langle 1_j \vert$.
    \subfiglabel{c}
    Measured distance function $D(t)$ from the equilibrium for initial states
    $\rho_1 = \vert 1_0 \rangle \langle 1_0 \vert$,
    $\rho_2 = 1/2\sum_{j=2}^{3}\vert 1_j \rangle \langle 1_j \vert$,
    $\rho_3 = 1/3\sum_{j=1}^{3}\vert 1_j \rangle \langle 1_j \vert$, and 
    $\rho_4 = 1/4\sum_{j=0}^{3}\vert 1_j \rangle \langle 1_j \vert$.
    Solid lines are numerical simulation results.
    $\rho_1$, the farthest from the equilibrium state, relaxes faster than the other three mixed states closer to the equilibrium state $\rho_E=1/7\sum_{j} \vert 1_j \rangle \langle 1_j \vert$, illustrating the \revise{Mpemba-effect-like dynamics}.
    }
\end{figure}

We finally explore the \revise{Mpemba-effect-like dynamics} from the dynamical behavior toward the equilibrium of the diffusive dynamics.
The so-called \revise{Mpemba-effect-like dynamics} is unveiled by quantitatively evaluating the relaxation dynamics~\cite{Longhi2024a},
which describes a scenario where a far-from-equilibrium state can relax toward equilibrium faster than a state closer to equilibrium~\cite{Bechhoefer2021}.
The classic example is that warmer water can sometimes freeze faster than colder water~\cite{Mpemba1969}.
Recently, quantum versions of this effect~\cite{Carollo2021} have been investigated in different quantum systems, including a quantum dot coupled with two reservoirs~\cite{Chatterjee2023}, a coherently driven trapped ion qubit coupled to a thermal Markovian bath~\cite{AharonyShapira2024}, and an array of trapped ions with power-law decaying interactions \cite{Joshi2024}.
In the single-excitation subspace of the 1D system where the dephasing rate is much larger than energy relaxation, any initial state has a non-vanishing projection onto the equilibrium state $\rho_E = 1/L\sum_{j} \vert 1_j \rangle \langle 1_j \vert$ spreading in all sites of the lattice, 
which leads to a relaxation dynamics toward $\rho_E$ in a long-time limit $Rt\gg 1$ with a diffusion rate $R=2 J^2/\Gamma$~\cite{Longhi2024}.
\revise{Mpemba-effect-like dynamics} with dephasing noise is evaluated by the observable distance function $D(t)$ for a given state $\rho(t)$ from the equilibrium state $\rho_E$ defined as
$D(t)=\operatorname{Tr}\left[\rho(t) \log\rho(t)\right]-\operatorname{Tr}\left[\rho_E(t) \log\rho_E(t)\right]$.
In the case of single-particle excitation,
it takes the following form in terms of the population $n_j(t)$~\cite{Longhi2024a}
\begin{equation}
D(t)=\log L + \sum_j n_j(t) \log n_j(t).
\end{equation}
To experimentally observe the \revise{Mpemba-effect-like dynamics}, a small system size is more favorable~\cite{Longhi2024a}.
As the system size $L$ increases, the uniformly distributed excitation of the equilibrium state has a smaller population at each site, and the relaxation dynamics become slower to reach equilibrium.

We now demonstrate the \revise{Mpemba-effect-like dynamics} in our 1D chain of $L=7$ qubits,
with the homogeneous coupling strengths $g_j=J\approx 2\pi \times 8.3$~MHz and $\Gamma/J\approx 3$ used to accelerate the diffusive dynamics toward equilibrium state $\rho_E$.
As shown in \rsubfig{fig4}{a}, when the system is initially prepared as a highly localized pure state $\rho_1 = \vert 1_{j=0} \rangle \langle 1_{j=0} \vert$ with the excitation at site $j=0$, it shows spreading of qubit population $n_j(t)$ in all sites when the evolution time $t$ is large enough.
In a similar scenario, we initialize the system as a mixed state $\rho_4 = 1/4\sum_{j=0}^{3} \vert 1_j \rangle \langle 1_j \vert$ by randomly applying the $\pi$ pulse on the sites $j=0,1,2,3$, which is closer to the stationary state $\rho_E$ and gives a temporal evolution in \rsubfig{fig4}{b}, showing a slower diffusion toward the missing occupation on the other sites.
Intuitively, the initial state with a closer distance to the equilibrium state is expected to diffuse faster and fill the missing occupation of other sites.
To perform a systematic study, we prepare two more initial states $\rho_2 = 1/2\sum_{j=2}^{3}\vert 1_j \rangle \langle 1_j \vert$,
$\rho_3 = 1/3\sum_{j=1}^{3}\vert 1_j \rangle \langle 1_j \vert$ between $\rho_1$ and $\rho_4$ in terms of distance to $\rho_E$, and observe their dynamics towards equilibrium similar to \rsubfigsA{fig4}{a}{b}, see \SeeSupply{} for details and numerical simulations.
The deviation between the measured population and corresponding numerical results is mainly limited by two-level system defects and readout errors.
By evaluating the distance function $D(t)$ from the measured population $n_j(t)$,
we present the results of $D(t)$ for different initial states in \rsubfig{fig4}{c}, which experimentally demonstrate that the farthest-from equilibrium state $\rho_1$ relaxes faster than the other three mixed states being closer to the equilibrium state $\rho_E$, clearly illustrating the \revise{Mpemba-effect-like dynamics}.

In this work, we experimentally demonstrate the diffusive dynamics assisted by controlled dephasing noise in a 1D array of superconducting qubits.
By properly arranging the coupling strengths between neighboring qubits, we implement a quasiperiodic lattice and demonstrate that
\revise{dephasing can give distinct dynamical behavior in this system}~\cite{Longhi2024}.
Moreover, we explore the \revise{Mpemba-effect-like dynamics} in this 1D model by preparing different excitation distributions in the qubit array and observing that more localized initial states relax to a uniformly distributed mixed state more rapidly under dephasing noise~\cite{Longhi2024a,Moroder2024}.
The experimental demonstration of these two counterexamples not only deepens our understanding of diffusive dynamics at the microscopic level, but also highlights the expanded potential of controlled dissipative processes in quantum simulation.
Utilizing the dissipative effect synthesized in such fully controlled quantum systems, 
further efforts can be made for the studies, including exotic non-equilibrium phases~\cite{Cai2014,Maghrebi2016}, reservoir computing approach~\cite{Angelatos2021}, and quantum thermodynamics~\cite{Talkner2020,Cech2023}.

See the supplementary material for detailed results on the device information, numerical simulations, and additional experiment results.

We thank Yucheng Wang, Xinchi Zhou, and Yuanzhen Chen for their helpful discussions.
This work was supported by the Science, Technology and Innovation Commission of Shenzhen Municipality (KQTD20210811090049034, RCBS20231211090824040, RCBS20231211090815032), the National Natural Science Foundation of China (12174178, 12204228, 12374474 and 123b2071), the Innovation Program for Quantum Science and Technology (2021ZD0301703), the Shenzhen-Hong Kong Cooperation Zone for Technology and Innovation (HZQB-KCZYB-2020050), and Guangdong Basic and Applied Basic Research Foundation (2024A1515011714, 2022A1515110615).

\vspace{0.15cm}
\subpart{AUTHOR DECLARATIONS}

\subpart{Conflict of Interest}

The authors have no conflicts to disclose.

\vspace{0.15cm}
\subpart{Author contributions}
\par \noindent
Y.L. and C.X. contributed equally to this work.
C.X. and Y.L. performed the experiments and theoretical analysis under the supervision of W.R. and Z.T.
Youpeng Z., W.R., and Z.T. supervised and coordinated the project.
All authors contributed to the discussion and preparation of the manuscript.

\vspace{0.15cm}
\subpart{Data availability}
\par \noindent
The data that support the findings of this study are available from the corresponding authors upon reasonable request.

\bibliographystyle{ref_style}

\end{document}


\title{Supplementary Material for ``Dephasing-assisted diffusive dynamics in superconducting quantum circuits"}
\maketitle

\setcounter{equation}{0}
\setcounter{figure}{0}
\setcounter{table}{0}

\newcommand{\subfiglabel}[1]{({#1})}
\newcommand{\rsubfiglabel}[1]{({#1})}
\newcommand{\rsubfig}[2]{Fig.~\textcolor{blueprl}{\ref{#1}}(#2)}
\newcommand{\rsubfigs}[3]{Fig.~\textcolor{blueprl}{\ref{#1}}(#2)-(#3)}
\newcommand{\rsubfigsA}[3]{Fig.~\textcolor{blueprl}{\ref{#1}}(#2) and (#3)}

\renewcommand{\theequation}{S\arabic{equation}}
\renewcommand{\thefigure}{S\arabic{figure}}
\renewcommand{\thetable}{S\arabic{table}}

\renewcommand{\figurename}{Fig.}
\def\figureautorefname{Fig.}

\newcommand{\gqq}[0]{g^{\mathrm{qq}}_j}
\newcommand{\gqcj}[0]{g^{\mathrm{qc}}_j}
\newcommand{\cplr}[0]{\mathrm{c}}
\newcommand{\wcp}[0]{\omega^{\mathrm{c}}}
\newcommand{\wcpj}[0]{\omega^{\mathrm{c}}_j}

\tableofcontents
\newpage

\section{Device and experimental setup}
\subsection{Device information}

\begin{figure}[htbp] 
  \centering
  \includegraphics[width=0.7\textwidth]{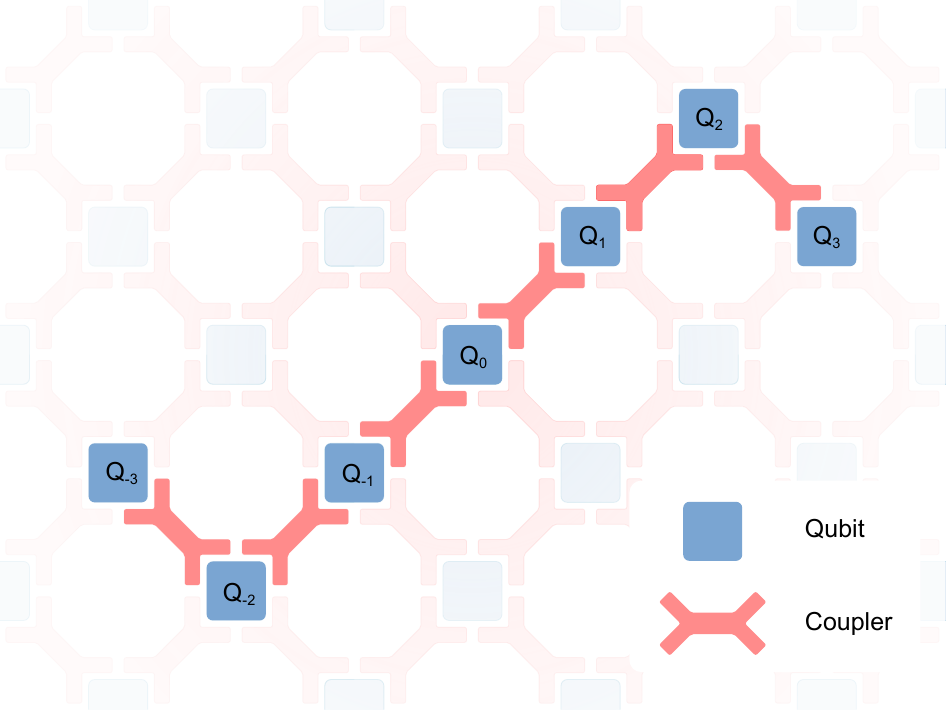}
  \caption{
  Schematic of the components selected from the superconducting quantum processor. The blue squares represent the qubits and the red bone-shaped patterns indicate the couplers between the qubits.
  }
  \label{Device}
\end{figure}

Our experiment is conducted on a quantum processor with $6\times 11$ qubits arranged in a grid array, where transmon-style tunable couplers~\cite{Xu2020} are utilized to couple neighboring qubits, see Ref.~\cite{Yang2024} for more details.
Each qubit is equipped with an asymmetric superconducting quantum interference device (SQUID), with frequencies ranging from about 3.7 GHz to 5.3 GHz.
The qubit readout frequencies span from 6.0 GHz to 6.5 GHz, and every six resonators share a Purcell filter. 
To experimentally realize the lattice model, we select a $7$-qubit array from the processor, labeled as $Q_j$ $(j = 3, 2, \dots, -3)$, as shown in \rfig{Device}. Characteristic parameters for these qubits are detailed in \rtab{Tab01}.

\begingroup
\begin{table}[!t]
    \centering
    \renewcommand{\arraystretch}{0.8}
    \begin{tabular}{cccccccc}
     \toprule
      & $Q_3$ & $Q_2$ & $Q_1$ & $Q_0$ & $Q_{-1}$ & $Q_{-2}$ &$Q_{-3}$\\
     \midrule
     $\omega_{j, \textrm{min}}/2\pi$ (GHz) & 3.79 & 3.71 & 3.96 & 4.00 & 3.95 & 3.91 & 3.90\\
     $\omega_{j, \textrm{max}}/2\pi$ (GHz) & 5.24 & 5.20 & 5.19 & 5.29 & 5.30 & 5.19 & 5.20\\
     $\omega_{j, \textrm{idle}}/2\pi$ (GHz) & 4.140 & 4.402 & 4.093 & 4.052 & 4.16 & 4.248 & 4.146 \\
     $\omega_{j, \textrm{r}}/2\pi$ (GHz) & 6.116 & 6.210 & 6.113 & 6.162 & 6.190 & 6.082 & 6.187 \\
     $T_{1j, \textrm{idle}}$ ($\mu$s) & 76.4 & 67.8 & 64.3 & 69.7 & 64.3 & 80.7 & 56.2 \\
     $T_{1j, \textrm{work}}$ ($\mu$s) & 76.2 & 61.3 & 46.7 & 73.0 & 45.9 & 71.9 & 54.8 \\
     $T_{2j, \phi}$ ($\mu$s) & 3.0 & 5.0 & 3.6 & 3.2 & 7.2 & 8.3 & 8.7\\
     $F_{0, j
     }$ & 0.986 & 0.976 & 0.975 & 0.974 & 0.969 & 0.977 & 0.984 \\
     $F_{1, j}$ & 0.973 & 0.943 & 0.959 & 0.930 & 0.929 & 0.952 & 0.961 \\
     \bottomrule
\end{tabular}
\caption{
\label{Tab01}
Qubits parameters. $\omega_{j, \textrm{min}}/2\pi$ and $\omega_{j, \textrm{max}}/2\pi$ are the minimum and maximum frequency of $Q_j$ respectively, whereas $\omega_{j, \textrm{idle}}/2\pi$ represents the idle frequency of $Q_j$. The relaxation times $T_{1j, \textrm{idle}}$ and $T_{1j, \textrm{work}}$ represent the coherence lifetimes of $Q_j$ at idle and operating frequencies, respectively. $T_{2j, \phi}$ denotes the dephasing time of $Q_j$. $F_{0, j}$ ($F_{1, j}$) indicates the measured fidelity of $Q_j$ in the state $|0\rangle$ ($|1\rangle$) when prepared in this state.}
\end{table}
\endgroup

\begin{figure}[!t]
  \centering
  \includegraphics[width=0.77\textwidth]{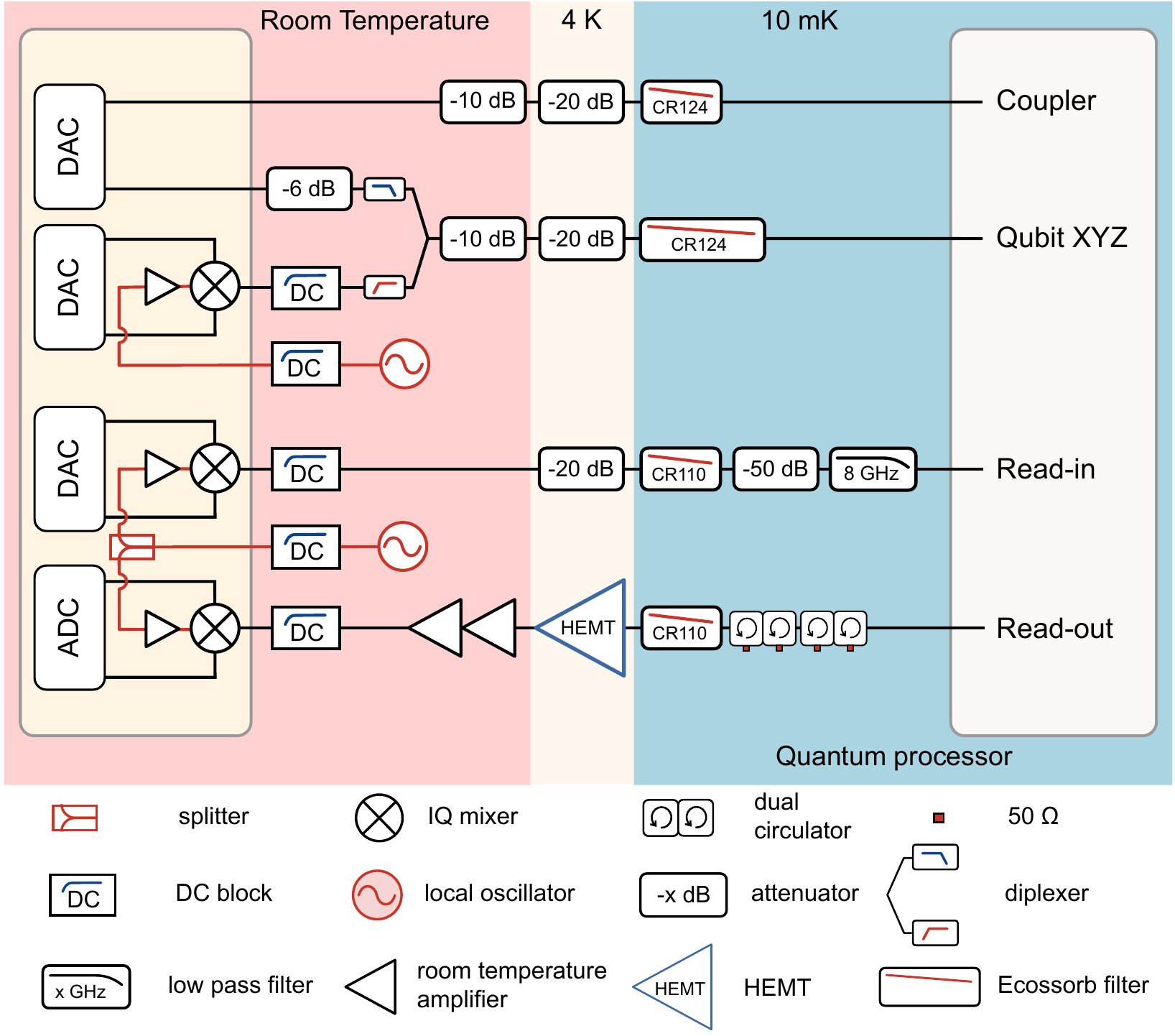}
  \caption{Schematic diagram of the experimental system and wiring information.}
  \label{wiring}
\end{figure}

\subsection{Experimental setup}

\rfig{wiring} illustrates the experiment's cryogenic wiring configuration and the room-temperature electronics. 
We employ custom-made digital-to-analog converter (DAC) and analog-to-digital converter (ADC) boards to control qubits and couplers. The baseband signals generated by the DAC boards are utilized to tune the frequencies of the qubits and couplers (i.e., Z bias). The DACs can generate microwave control signals through IQ mixing after up-conversion, which are used for XY control and qubit state readout.
A diplexer combines the XY control signal and the Z bias signal for each qubit.
Attenuators and filters are properly installed in the control lines to mitigate thermal noise. The readout output signal is amplified by a high electron mobility transistor (HEMT) low noise amplifier from Low Noise Factory, followed by further amplification from an additional amplifier at room-temperature, then digitized by the ADC board and demodulated for qubit state discrimination.

\section{Dephasing noise generation on superconducting qubits}

\begin{figure}[!htbp]
  \centering
  \includegraphics[width=0.9\textwidth]{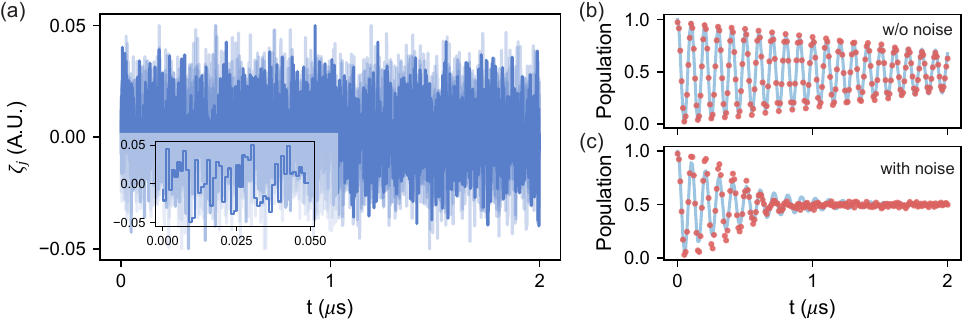}
  \caption{
  Illustration of the dephasing noise generation.
  \subfiglabel{a}
  Example of a random noise sequence $\zeta_j$ consisting of $N=200$ trajectories. Each trace represents a trajectory, where the inset is a zoomed in view of the first 50~ns of a trajectory, showing the rectangular pulses of each time step.
  \subfiglabel{b}-\subfiglabel{c}
    Example of Ramsey interference measurements without \subfiglabel{b} and with \subfiglabel{c} applied dephasing noise.
  }\label{SI_noise_ramsey}
\end{figure}

In this work, we follow the method discussed in Ref.~\cite{Averin2016} to inject artificial dephasing noise into superconducting qubits.
We apply a series of pulse sequences in the experiments and regard the measured outputs as an ensemble of trajectories,
in which the average population gives an effective evolution of the simulated quantum system with engineered decoherence processes.
Each pulse sequence consists of successive square pulses and can be specifically designed to yield the desired noise spectrum.
As illustrated in Ref.~\cite{Averin2016}, the noise spectral density $S(\omega)$ is first discretized into the series $S(\omega_m)$,
where $\omega$ is the angular frequency, $\omega_m = 2\pi m/(Nk\tau_0)$ ($m=0,1,2,\cdots ,Nk/2$), $N$ is the number of pulse sequences,
$k$ is the maximum number of successive rectangle pulses in each sequence, and $\tau_0=1$~ns is the time step of these rectangle pulses. 
In the case of Gaussian white noise used in our experiment, $S(\omega)$ is simply a constant whose amplitude determines the noise strength.
We then multiply $\sqrt{S (\omega_m)}$ by a phase factor $\theta_m$ randomly chosen from $0$ to $2\pi$, 
and perform the inverse Fourier transform of the frequency series 
$\{\sqrt{S(\omega_m)}\exp(i\theta_m)$  $,m=0,1,\cdots Nk/2\}$ 
to generate a discrete white noise time series $\xi(t)$ consisting of $Nk$ numbers.
The time series $\xi(t)$ is then divided into $N$ trajectories, where each trajectory is transformed to the amplitudes of Z bias pulses experimentally generated by the DAC modules, as shown in \rsubfig{SI_noise_ramsey}{a}.
The generated pulses are finally applied to the superconducting qubits through the control line to modulate their frequencies. 
As an example, the measured results of Ramsey interference for one qubit, both without and with controlled noise, are shown in \rsubfigs{SI_noise_ramsey}{b}{c}, respectively.
\begin{reviseAPL}
    For the delta-correlated (white) Gaussian noise with zero mean and amplitude $\Gamma$ \cite{Schmolke2022}, where 
    \begin{equation}
        \langle \xi (t) \xi (t')\rangle=\Gamma \delta(t-t').
    \end{equation}
    The relationship between the spectral density $S(\omega)$ of the applied noise and $\Gamma$ can be expressed as \cite{Averin2016}
    \begin{equation}
        \langle \xi (0) \xi (t)\rangle = \int \frac{\mathrm{d}\omega}{2\pi} S(\omega) e^{-i\omega t} = \Gamma \delta(t).
    \end{equation}
    The noise applied on qubit $j$ can be described by $\mathcal{L}_j= \sqrt{\Gamma_j} \sigma^{+}_j\sigma^{-}_j$, where $\Gamma_j$ is also the effective dephasing rate generated by applied noise $\xi_j(t)$ with $\Gamma_j=\Gamma$.
    The system dynamics can be described by the corresponding master equation~\cite{Schmolke2022}:
    \begin{equation}
    \frac{\mathrm{d} \rho(t) }{\mathrm{d}t} = - \frac{i}{\hbar} [H,\rho(t)] + 
    \sum_j \left( \mathcal{L}_j\rho \mathcal{L}_j - \frac{1}{2} \left\{\mathcal{L}_j^\dagger \mathcal{L}_j, \rho \right\}\right).
    \end{equation}
\end{reviseAPL}

In the main text, the population dynamics shown in Fig. 2(c), Fig. 3(d) and (f) is averaged from $N=50$ trajectories,
and the population shown in Fig. 4 is averaged from $N=500$ trajectories to evaluate distance function from the equilibrium.

\section{Detailed information for tunable couplers}\label{sec:coupler}

Instead of simply regarding the tunable couplers as tools for modulating coupling strengths, we can also take the couplers into account for the full Hamiltonian
\begin{equation}\label{eq:hami_full_cp}
H/\hbar=
\sum_{j=-l}^{l} \omega_j  \sigma_{j}^{+}\sigma_{j}^{-} +
\sum_{j=-l+1}^{l} 
\wcpj \sigma_{j,\cplr}^{+}\sigma_{j,\cplr}^{-}
+ \sum_{j=-l+1}^{l} \left(
\gqq \sigma_{j}^{+}\sigma_{j-1}^{-}+
\gqcj \sigma_{j}^{+}\sigma_{j,\cplr}^{-}+ \gqcj \sigma_{j-1}^{+}\sigma_{j,\cplr}^{-}+
\mathrm{H.c.}
\right)
\end{equation}
where $\sigma^{+}_{j}$ ($\sigma_{j}^{-}$) represents the raising (lowering) operator for the qubit $Q_j$, 
$\sigma^{+}_{j,\cplr}$ ($\sigma_{j,\cplr}^{-}$) represents the raising (lowering) operator for the coupler $C_j$ between the qubits $Q_j$ and $Q_{j-1}$, 
$\omega_j/2\pi$ ($\wcpj/2\pi$) is the frequency of qubit $Q_j$ (coupler $C_j$),
$\gqq$ ($\gqcj$) is the capacitive coupling between the neighboring qubits $Q_j$ and $Q_{j-1}$ (qubit $Q_j$ and coupler $C_j$).
Since the effective qubit-qubit coupling is partly mediated by the couplers, this full model enables us to consider the impact of tunable couplers on qubit dynamics.

In our device, the qubit operating frequencies are set to $\omega_j/2\pi=\omega_q/2\pi \approx 4.2$ GHz when the qubits are tuned to resonance.
The coupling parameters are estimated to be $\gqq/2\pi \approx 8$ MHz at the frequency $\omega_q/2\pi$,
and $\gqcj/2\pi \approx 110~\mathrm{MHz} \times \sqrt{\wcpj/\omega_q}$ at a given coupler frequency $\wcpj$,
and the maximum frequencies of tunable couplers are around 7.2 GHz~\cite{Yang2024}.
We tune the coupler frequencies $\wcpj/2\pi \approx 5.07$ GHz to modulate the effective coupling strengths between adjacent qubits $Q_j$ and $Q_{j-1}$ as $g_j = J \approx 2\pi \times 8.3$ MHz, which emulates the effective Hamiltonian:
 \begin{equation}\label{hami}
    H_{\mathrm{eff}}/\hbar=\sum_{j=-l+1}^{l}g_j(\sigma_j^{+}\sigma_{j-1}^{-}+\sigma_j^{-}\sigma_{j-1}^{+}),
\end{equation}
where $g_j$ is the effective coupling strength between adjacent qubits $Q_j$ and $Q_{j-1}$.
For other configurations of different $\kappa=B/A$ as used in Fig.~3 of the main text, the coupler frequencies are given in \rtab{Tab_Wcp}, and the effective qubit-qubit coupling strengths are given in \rtab{Tab_geff}.

\begin{table}[!htbp]
	\centering
			\begin{tabular}{cccccccccc}
				\toprule
				\quad & $\kappa$ & & $\wcp_3/2\pi$ & $\wcp_2/2\pi$ & $\wcp_1/2\pi$ & $\wcp_0/2\pi$ & $\wcp_{-1}/2\pi$ & $\wcp_{-2}/2\pi$ & (GHz)\\
				\midrule
                \quad &0.3&&5.12 &5.21 &5.37 &5.07 &5.37 & 5.21\\
				\quad &0.5&&5.16 &5.28 &5.58 &5.07 &5.58 & 5.28\\
				\quad &0.7&&5.17 &5.35 &5.79 &5.07 &5.79 & 5.35\\
				\quad &0.9&&5.19 &5.41 &6.01 &5.07 &6.01 & 5.41 \\
				\bottomrule
			\end{tabular}
		\caption{\label{Tab_Wcp}
			{Values of the operating frequencies for the couplers at different $\kappa$.}}
\end{table}

\begin{table}[!htbp]
	\centering
	\begin{tabular}{cccccccccc}
				\toprule
				\quad & $\kappa$ & & $g_3/2\pi$ & $g_2/2\pi$ & $g_1/2\pi$ & $g_0/2\pi$ & $g_{-1}/2\pi$ & $g_{-2}/2\pi$ & (MHz)\\
				\midrule
				\quad &0.3&&7.6&6.6&5.0&8.3&5.0&6.6\\
				\quad &0.5&&7.2&5.8&3.5&8.3&3.5&5.8\\
				\quad &0.7&&7.0&5.2&2.4&8.3&2.4&5.2\\
				\quad &0.9&&6.8&4.7&1.5&8.3&1.5&4.7\\
				\bottomrule
			\end{tabular}
		\caption{\label{Tab_geff}
			{Values of effective qubit-qubit coupling strengths at different $\kappa$.}}
	\end{table}

\begin{figure}[!htbp]
    \centering
	\includegraphics[width=0.75\textwidth]{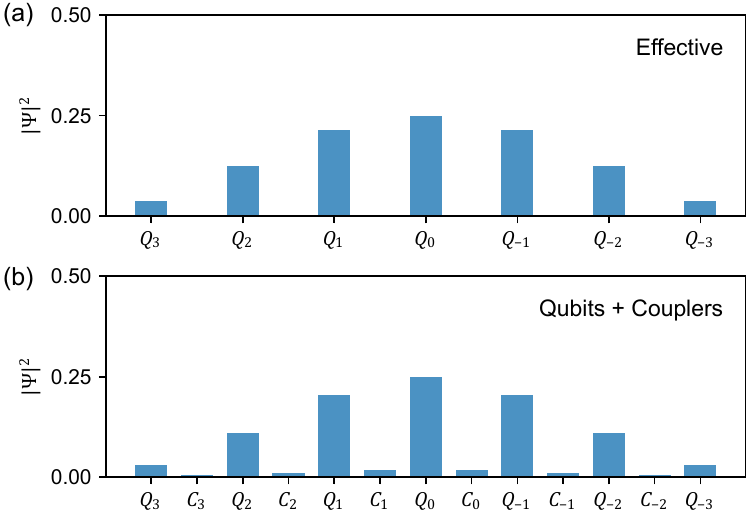}
    \caption{
    The calculated eigenmodes for the effective 1D chain and the entire qubits-couplers system.
    \subfiglabel{a}
    The distribution of lowest eigenmode for the 1D chain with single excitation.
    \subfiglabel{b}
    The corresponding distribution of eigenmode for the entire qubits-couplers system,
    which gives leakage population on the couplers below $10^{-1}$.
    The qubits (couplers) are labeled as $Q_j$ ($C_j$).
    }\label{EigPlot_eg1}
\end{figure}

\begin{figure}[!htbp]
    \centering
    \includegraphics[width=0.7\textwidth]{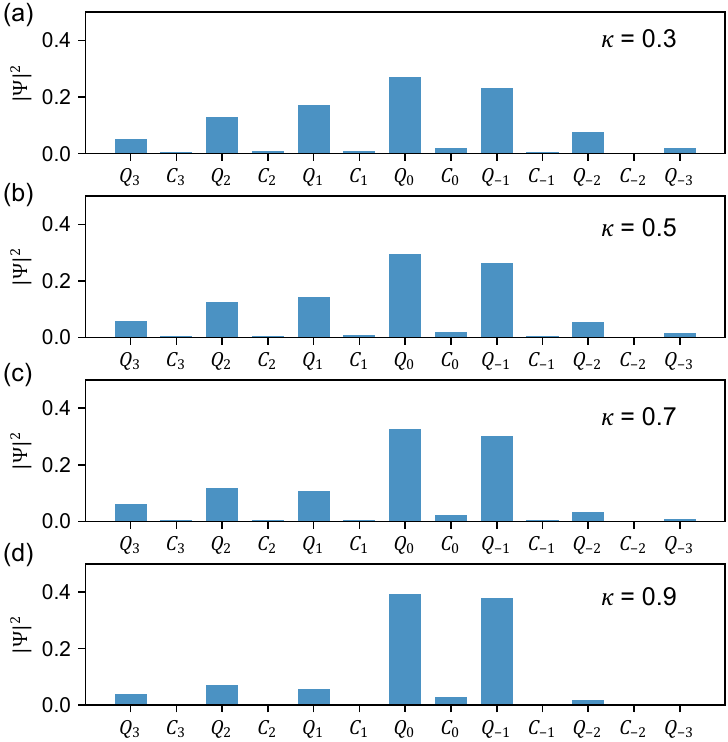}
    \caption{
        The distribution of lowest eigenmode with single excitation for the entire qubits-couplers  system, where the operating frequencies for the couplers are used from the parameters listed in \rtab{Tab_Wcp} for 
        \subfiglabel{a} $\kappa=0.3$,
        \subfiglabel{b} $\kappa=0.5$,
        \subfiglabel{c} $\kappa=0.7$,
        \subfiglabel{d} $\kappa=0.9$.
    }\label{EigPlots_eg2}
\end{figure}

To further consider the impact of tunable couplers, we calculate the eigenmodes of the effective 1D chain and the full qubits-couplers system, as shown in \rsubfigsA{EigPlot_eg1}{a}{b} respectively.
\rfig{EigPlots_eg2} shows the eigenmodes distributions for the full qubits-couplers system under different configurations of the couplers' operating frequencies listed in \rtab{Tab_Wcp}.
When the couplers are close in frequency to the qubits, the leakage population on the couplers exists in the entire qubit-coupler system with tolerable deviations on the subsystem of qubits used in the simulation of the effective 1D chain, 
as shown in the distribution of eigenmodes, 
which can be further suppressed by decreasing the effective coupling strengths when required.
Considering the impact of tunable couplers is more applicable to describe practical scenarios in the experiments.
In subsequent sections, we also present numerical simulations for the qubit dynamics when considering the tunable couplers detailed in this section.

\section{Diffusive spreading dynamics with dephasing noise}

\begin{figure}[!t]
  \centering
  \includegraphics[width=0.8\textwidth]{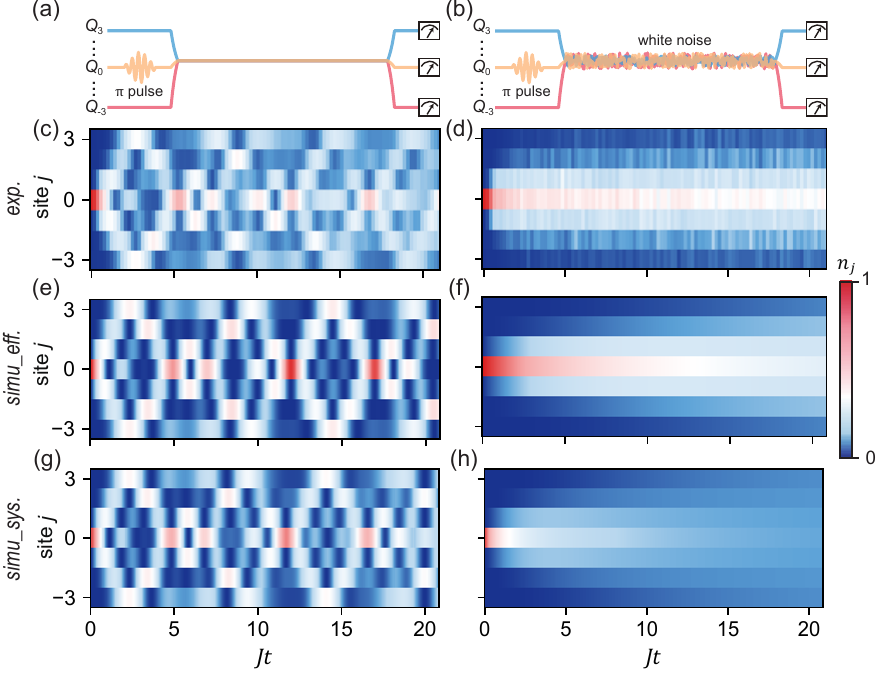}
  \caption{
  Supplementary numerical simulations for diffusive spreading dynamics.
  \subfiglabel{a}-\subfiglabel{b}
  Illustration of the pulse sequence without \rsubfiglabel{a} and with \rsubfiglabel{b} noise.
  \subfiglabel{c}-\subfiglabel{d}
  The experimental data for the population dynamics $\langle n_j \rangle$ without \rsubfiglabel{c} and with \rsubfiglabel{d} noise.
  \subfiglabel{e}-\subfiglabel{f}
  The numerical results for the effective 1D chain described by \req{eq:hami_full_cp} corresponding to \rsubfiglabel{c} and \rsubfiglabel{d}, respectively.
  \subfiglabel{g}-\subfiglabel{h}
  The numerical results for the qubits-couplers system described by \req{eq:hami_full_cp} and detailed in \rsec{sec:coupler}.
}\label{maintexfig2}
\end{figure}

In our experiment, we consider a 1D chain of tunably coupled qubits with the same frequencies, as discussed in \rsec{sec:coupler}.
When noise is applied as illustrated in \rfig{maintexfig2}(b), 
the system evolution is governed by the Lindblad master equation:
\begin{equation}\label{eq:lindblad}
	\frac{d\rho}{dt} = -\frac{i}{\hbar} \left[H, \rho\right] + 
	\sum_{j} \left(\mathcal{L}_j \rho \mathcal{L}_j^\dagger - \frac{1}{2} \left\{\mathcal{L}_j^\dagger \mathcal{L}_j ,\rho \right\} \right),
\end{equation}
where $\rho$ is the density matrix, $\mathcal{L}_j=\sqrt{\Gamma_j}\sigma_{j}^{+}\sigma_{j}^{-}$ denotes the dephasing collapse operator, and $\Gamma_j$ is the effective dephasing rate generated by applied noise $\xi_j(t)$ with $\Gamma_j=\Gamma$.
In numerical simulation, we calculate the system evolution using the Lindblad master equation with QuTiP python library~\cite{Johansson2012}.

We first consider that the coupling strength $g_j/2\pi$ between qubits is uniformly 8.3 MHz.
The pulse sequence for implementing this model is illustrated in \rsubfig{maintexfig2}{a},
in which a $\pi$-pulse is firstly applied to the central qubit.
All of the qubits are tuned to resonance, and the population $\langle n_j\rangle$ is finally measured.
The experimental data are given in \rsubfigs{maintexfig2}{c}{d}, and the corresponding numerical simulations of the effective 1D chain described by \req{hami} are presented in \rsubfigs{maintexfig2}{e}{f}.
\rsubfigsA{maintexfig2}{c,e}{d,f} present the results of population dynamics $n_j=\langle 1_j \vert \rho(t)\vert 1_j \rangle$ without and with applied noise, respectively, where the effective dephasing rate $\Gamma=30J$ is used in the latter case.
In the numerical simulations for the qubits-couplers system described by \req{eq:hami_full_cp} shown in \rsubfigs{maintexfig2}{g}{h} and the following numerical results for the qubits-couplers system, we consider the relaxation of both qubits and tunable couplers by using the parameters including the detailed coupling parameters discussed in \rsec{sec:coupler}, the qubit relaxation time given in \rtab{Tab01}, and the couplers relaxation time as $10~\mu\mathrm{s}$~\cite{Yang2024}.
When the tunable couplers are considered, the dephasing paramters are chosen to keep the similar diffusive dynamics given in the effective Hamiltonian.

\begin{figure}[!htbp]
  \centering
  \includegraphics[width=0.75\textwidth]{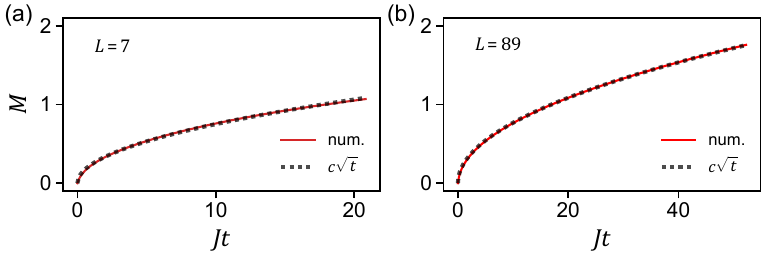}
  \caption{
  Numerical simulation of the integrated moment $M(t)$ for different system sizes $L$.
  \subfiglabel{a}-\subfiglabel{b}
  Calculated $M(t)$ as a function of normalized time $Jt$ given by the red solid line for \rsubfiglabel{a} $L=7$ and \rsubfiglabel{b} $L=89$,
  where the diffusive scaling $c\sqrt{t}$ is shown by black dashed lines.
  }\label{L7andLlarge}
\end{figure}

\begin{figure}[!htbp]
  \centering
  \includegraphics[width=1\textwidth]{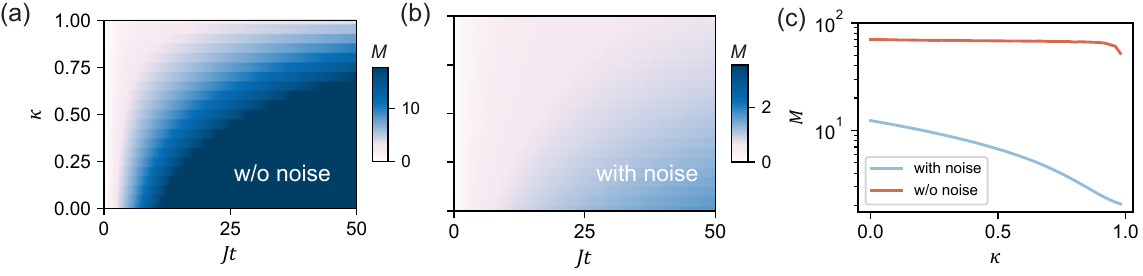}
  \caption{
  Numerical simulation for the off-diagonal Aubry-Andr\'e model. 
  \subfiglabel{a}-\subfiglabel{b}
  The integrated moment $M$ as a function of $Jt$ and $\kappa$ without \rsubfiglabel{a} and with \rsubfiglabel{b} applied noise during the early time dynamics. 
  \subfiglabel{c}
  The integrated moment $M$ versus $\kappa$ for $t=50~\mu$s with $J/2\pi\approx 8.3~\mathrm{MHz}$ as an example of long evolution time,
  where the red (blue) line represents the scenario without (with) noise.
  The red line shows a slightly decreased response for the observable $M$ near the boundary of $\kappa < 1$ due to the finite-size effects on system size and evolution time.
  The system size is $L=89$ in this simulation.
  }\label{M_vs_kappa}
\end{figure}

Comparing the scenarios without and with controlled noise, the population dynamics exhibit ballistic and diffusive spreading, respectively. 
As discussed in the main text, we characterize these spreading dynamics using the integrated moment $M(t)=(1/t)\int_{0}^{t}[W(\tau)-W(0)]^{1/2}d\tau$,
where $W(\tau)=\sum_{j}{|j-j_{0}|^2} n_{j}(\tau)$ is effective second moment.
Fig. 2(d) in the main text gives $M(t)$ as a function of normalized time $Jt$, which shows a characteristic scaling of $M(t)\sim t^{0.5}$ for diffusive spreading.
Here, we present supplementary numerical simulation in \rsubfigsA{L7andLlarge}{a}{b},
showing $M(t)$ versus $Jt$ for systems with $L=7$ and $L=89$ sites.
The red lines correspond to numerical results calculated from \req{hami} and \req{eq:lindblad}, while the gray dash lines represent the scaling function $c\sqrt{t}$, with a constant $c$.
We note that $M(t)$ for both the system size $L=7$ and $L=89$ fits the scaling function $M(t)\sim t^{0.5}$.

\begin{figure}[!htbp]
  \centering
  \includegraphics[width=0.95\textwidth]{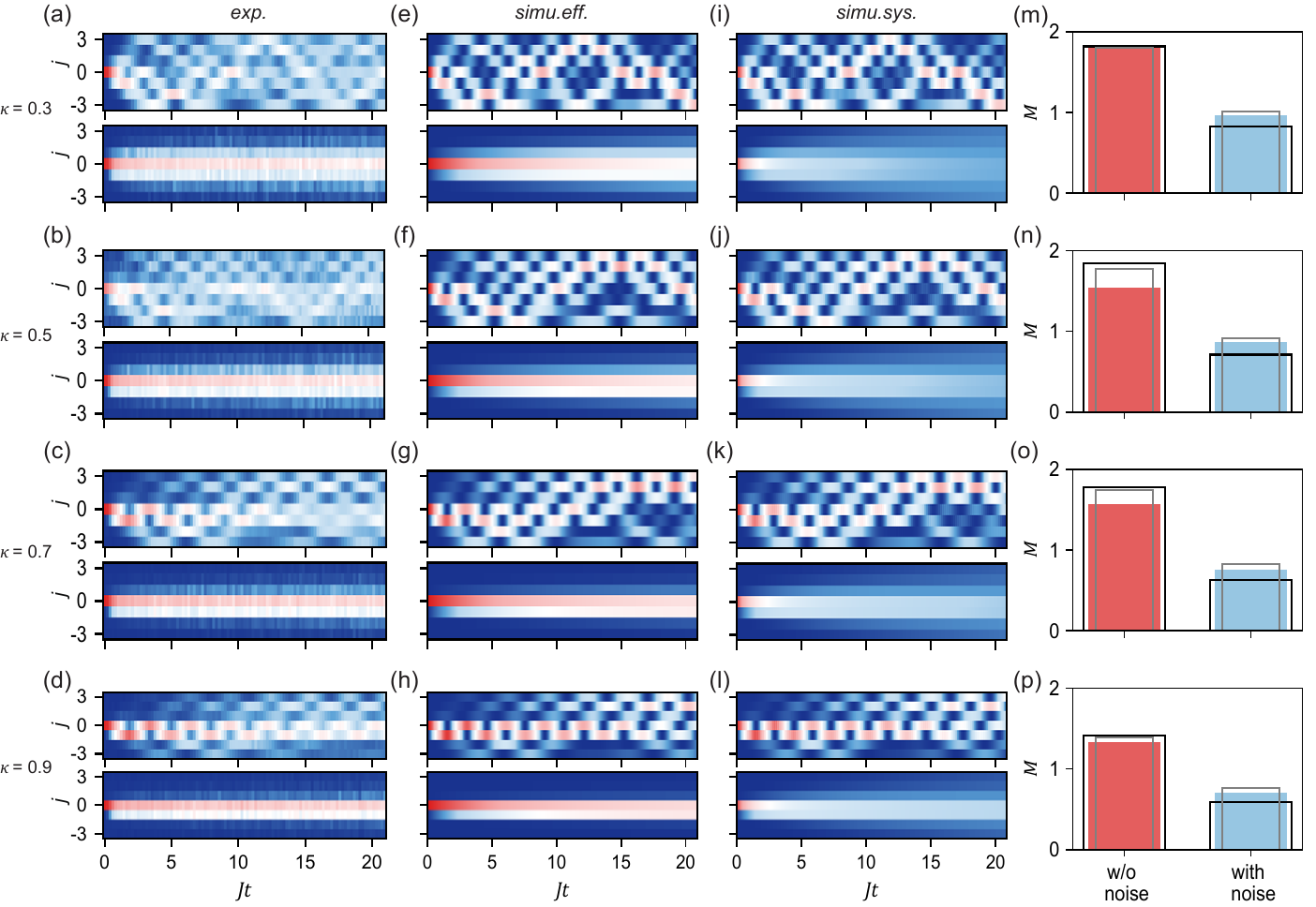}
  \caption{Experimental and numerical results of population $\langle n_j \rangle$ and integrated moment $M$ in the system with quasiperiodic coupling strengths.
  \subfiglabel{a}-\subfiglabel{d}
  The experimental data for the population $n_j$ without (top) and with (bottom) noise, 
  where the ratio $\kappa = 0.3,0.5,0.7,0.9$ \rsubfiglabel{a,b,c,d}.
  \subfiglabel{e}-\subfiglabel{h}
  The corresponding numerical simulations of the effective 1D chain described by \req{hami} for panels \rsubfiglabel{a}-\rsubfiglabel{d}.
  \subfiglabel{i}-\subfiglabel{l}
  The corresponding numerical simulations of the qubits-couplers system described by \req{eq:hami_full_cp}.
  \subfiglabel{m}-\subfiglabel{p}
  The integrated moment $M$ evaluated from the qubit dynamics.
  The solid bars represents the experimental results,
  and the black (gray) frames represent the numerical results for the effective 1D chain (qubits-couplers system).
  The effective dephasing rate $\Gamma/J=30$ is used here.
  }\label{maintexfig3}
\end{figure}

Then we consider the off-diagonal Aubry-Andr\'e model with coupling strengths $g_j =A+B\cos(2\pi \alpha j+\theta)$.
For this model, we configure the system parameters as $A+B=J \approx 2\pi \times 8.3$ MHz,
irrational frequency $\alpha=\left(\sqrt{5}-1 \right)/2$, and phase offset $\theta=0$.
%
The behavior of the integrated moment $M(t)$ for different values of $\kappa$ is depicted in \rfig{M_vs_kappa}.
The numerical results in the absence of controlled noise are shown in \rsubfig{M_vs_kappa}{a}, while the scenarios with controlled noise are presented in \rsubfig{M_vs_kappa}{b}.
\rsubfig{M_vs_kappa}{c} illustrates the accumulated $M(t)$ as a function of $\kappa$, which shows that
$M(t)$ without dephasing noise is an order of magnitude larger than that with noise.
Due to the finite-size effects on system size and evolution time,
the system without noise in the extended phase ($\kappa <1$) shows a decreased response for the observable $M(t)$ near the boundary of $\kappa < 1$.

\begin{figure}[!htbp]
  \centering
  \includegraphics[width=0.7\textwidth]{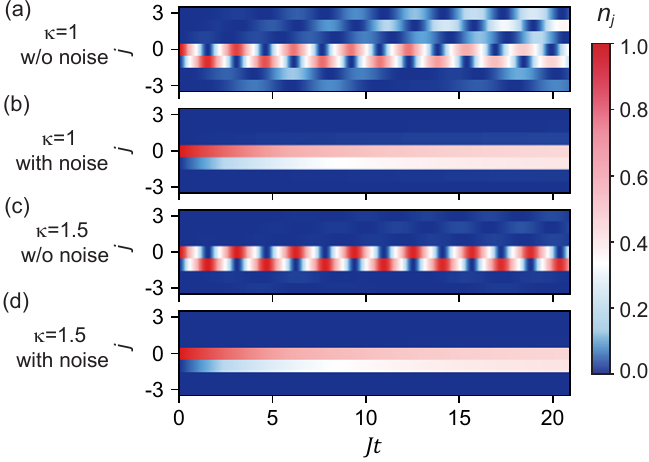}
 \begin{reviseAPL}
 \caption{
     Supplementary numerical results of population $n_j$ for 
 \subfiglabel{a}-\subfiglabel{b} $\kappa=1$ without \rsubfiglabel{a} and with \rsubfiglabel{b} noise,
 \subfiglabel{c}-\subfiglabel{d} $\kappa=1.5$ without \rsubfiglabel{c} and with noise \rsubfiglabel{d}.
 }\label{SI_simu_kappa_larger_1}
 \end{reviseAPL}
\end{figure}

As supplementary data for Fig. 3 in the main text,
we examine the population dynamics of the qubit array at coupling ratios $\kappa=0.3,0.5,0.7$ and $0.9$, as presented in \rsubfigs{maintexfig3}{a}{d}.
Each upper (lower) panel of \rsubfigs{maintexfig3}{a}{l} gives the results without (with) applied noise.
The numerical results of the effective 1D chain described in \req{hami} corresponding to these experimental results are presented in \rsubfigs{maintexfig3}{e}{h};
the numerical results of the qubits-couplers system described in \req{eq:hami_full_cp} are presented in \rsubfigs{maintexfig3}{i}{l}.
The integrated moments $M(t)$ accumulated during the evolution are given in \rsubfigs{maintexfig3}{m}{p},
where red and blue solid bars represent results without and with controlled noise, respectively,
while the black (gray) frames denote the numerical results for the effective 1D chain (qubits-couplers system) corresponding to \rsubfigs{maintexfig3}{e}{h} (\rsubfigs{maintexfig3}{i}{l}).
We note that the deviation between the measured population and corresponding numerical results is mainly limited by two-level system defects and readout errors, especially for the small values of population.
\revAPL{
Additionally, we present the numerical results of the population size $n_j$ for $\kappa\ge 1$, as displayed in \rsubfigs{SI_simu_kappa_larger_1}{a}{d},
which shows a similar diffusive dynamics with noise and can be further explored in future work.}

\revAPL{
\section{Mpemba-effect-like dynamics with dephasing noise}
}

\begin{figure}[!htbp]
  \centering
  \includegraphics[width=1\textwidth]{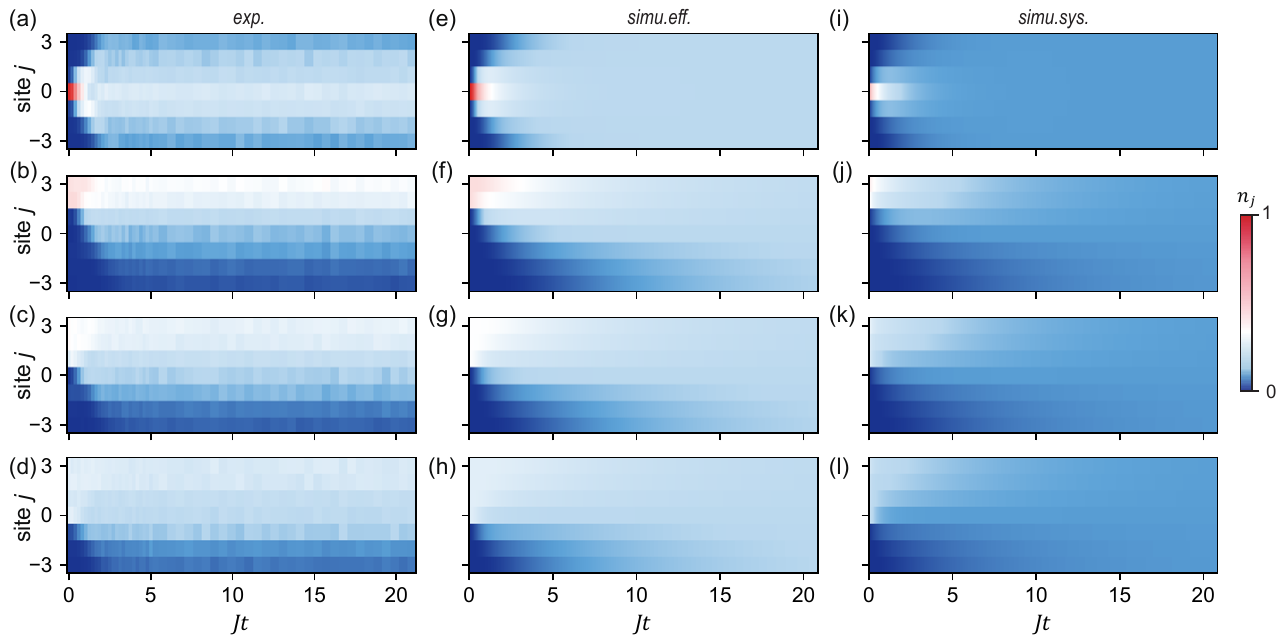}
  \reviseText{
  \caption{
  Supplementary data and numerical simulation for Mpemba effect with dephasing noise. 
  \subfiglabel{a}-\subfiglabel{d}
  The experimental data showing the evolution of $\langle n_j\rangle$ under dephasing noise,
  starting from different initial states: $\rho_1=|1_0\rangle\langle 1_0|$ (same data as Fig. 4(a) in the main text),
  $\rho_2=1/2\sum_{j=2}^{3}|1_j\rangle\langle 1_j|$,
  $\rho_3=1/3\sum_{j=1}^{3}|1_j\rangle\langle 1_j|$ and
  $\rho_4=1/4\sum_{j=0}^{3}|1_j\rangle\langle 1_j|$ (same data as Fig. 4(b) in the main text).
  \subfiglabel{e}-\subfiglabel{h}
  Corresponding numerical results of the effective 1D chain for panels \rsubfiglabel{a}-\rsubfiglabel{d}, respectively.
  \subfiglabel{i}-\subfiglabel{l}
  Corresponding numerical results of the qubits-couplers system.
  }\label{SI_Mpemba}
}
\end{figure}

In this section, we present numerical results that complement the experimental data in Fig. 4 of the main text, see \rfig{SI_Mpemba}.
\reviseText{
The numerical results of the effective 1D chain described in \req{hami} corresponding to these experimental results are presented in \rsubfigs{maintexfig3}{e}{h};
the numerical results of the qubits-couplers system described in \req{eq:hami_full_cp} are presented in \rsubfigs{maintexfig3}{i}{l}.
}
Both the experimental data and numerical simulations demonstrate that the initial state $\rho_1=|1_0\rangle\langle 1_0|$,
the most localized among the four initial states, reaches equilibrium (an utterly mixed state) faster than the other states.

\begin{figure*}[!htbp]
  \centering
  \includegraphics[width=0.5\textwidth]{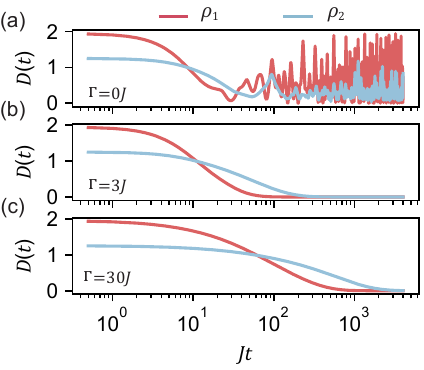}
  \caption{
  Numerical results of the distance function $D(t)$ to equilibrium for different dephasing rates $\Gamma$. 
  \subfiglabel{a}
  The numerical results of $D(t)$ with initial state $\rho_1=|1_0\rangle\langle 1_0|$ and $\rho_2=1/2\sum_{j=2}^3|1_j\rangle\langle 1_j|$ without controlled noise.
  \subfiglabel{b}-\subfiglabel{c}
  The numerical results of $D(t)$ with the same initial states $\rho_1$ and $\rho_2$ for effective dephasing rate \rsubfiglabel{b} $\Gamma/J=3$ and \rsubfiglabel{c} $\Gamma/J=30$.
  }\label{SI_Mpemba_2}
\end{figure*}

In Fig. 4(c) of the main text, the measured distances for different initial states with a dephasing rate $\Gamma/J=3$ are illustrated. 
Here we present numerical simulations at various dephasing rates in \rfig{SI_Mpemba_2}.
When $\Gamma/J=0$, the states evolve under coherent evolution which cannot reach equilibrium, as shown in \rsubfig{SI_Mpemba_2}{a}.
When the dephasing is involved, the system can relax toward equilibrium, and shows a faster relaxation dynamics for a smaller value of $\Gamma/J$.
As shown in \rsubfigsA{SI_Mpemba_2}{b}{c}, the distance function shows a similar behavior for $\Gamma/J=3$ and $\Gamma/J=30$, while the former case gives faster relaxation dynamics toward equilibrium, which is more favorable in experimental realization.

\clearpage